\documentclass[%
 reprint,
superscriptaddress,
 amsmath,amssymb,
 aps,
prstab,
]{revtex4-2}

\makeatletter
\def\bibsection{%
   \par
   \begingroup
    \baselineskip26\p@
    \bib@device{\hsize}{72\p@}%
   \endgroup
   \nobreak\@nobreaktrue
   \addvspace{19\p@}%
  }%
\makeatother

\usepackage{graphicx}
\usepackage{dcolumn}
\usepackage{bm}
\usepackage{siunitx}
\usepackage{hyperref}
\hypersetup{
	colorlinks=true,
	linkcolor=blue,
	filecolor=blue,      
	urlcolor=blue,
	citecolor=blue,
}

\begin{document}

\newcommand{\waket}{\mbox{Wake-T}}
\newcommand{\fbpic}{FBPIC}


\title{Bayesian optimization of laser-plasma accelerators \\assisted by reduced physical models}

\author{A. Ferran Pousa}
\email{angel.ferran.pousa@desy.de}
\affiliation{Deutsches Elektronen-Synchrotron DESY, Notkestr. 85, 22607 Hamburg, Germany }

\author{S. Jalas}
\affiliation{Department of Physics Universit\"{a}t Hamburg, Luruper Chaussee 149, 22761 Hamburg, Germany}

\author{M. Kirchen}
\affiliation{Deutsches Elektronen-Synchrotron DESY, Notkestr. 85, 22607 Hamburg, Germany }

\author{A. Martinez de la Ossa}
\affiliation{Deutsches Elektronen-Synchrotron DESY, Notkestr. 85, 22607 Hamburg, Germany }

\author{M. Th\'{e}venet}
\affiliation{Deutsches Elektronen-Synchrotron DESY, Notkestr. 85, 22607 Hamburg, Germany }
    
\author{\\S. Hudson}
\affiliation{Argonne National Laboratory, Lemont, IL 60439, USA }

\author{J. Larson}
\affiliation{Argonne National Laboratory, Lemont, IL 60439, USA }

\author{A. Huebl}
\affiliation{Lawrence Berkeley National Laboratory, Berkeley, CA 94720, USA }

\author{J.-L. Vay}
\affiliation{Lawrence Berkeley National Laboratory, Berkeley, CA 94720, USA }

\author{R. Lehe}
\affiliation{Lawrence Berkeley National Laboratory, Berkeley, CA 94720, USA }

\date{\today}

\begin{abstract}

Particle-in-cell simulations are among the most essential tools for the modeling and optimization of laser-plasma accelerators, since they reproduce the physics from first principles.
However, the high computational cost associated with them can severely limit the scope of parameter and design optimization studies.
Here, we show that a multitask Bayesian optimization algorithm can be used to mitigate the need for such high-fidelity simulations by incorporating information from inexpensive evaluations of reduced physical models.
In a proof-of-principle study, where a high-fidelity optimization with FBPIC is assisted by reduced-model simulations with Wake-T, the algorithm demonstrates an order-of-magnitude speedup.
This opens a path for the cost-effective optimization of laser-plasma accelerators in large parameter spaces, an important step towards fulfilling the high beam quality requirements of future applications.


\end{abstract}

\maketitle

\section{INTRODUCTION}

Laser-plasma accelerators (LPAs) make use of a plasma medium to transform the energy of a laser pulse into large longitudinal electric fields capable of accelerating particles to high energies in a short distance~\cite{PhysRevLett.43.267}.
This process depends on a complex interplay of non-linear physical phenomena that determine the final performance of the accelerator.
The laser-plasma interaction (manifested as laser self-focusing, dephasing, and depletion~\cite{RevModPhys.81.1229}), the injection of electrons into the plasma wake~\cite{modena1995electron,bulanov1997transverse, PhysRevE.58.R5257, PhysRevLett.76.2073,PhysRevLett.104.025003,PhysRevLett.104.025004,Faure2006,Gonsalves2011, PhysRevLett.110.185006}, the beam-plasma interaction (especially beam loading~\cite{vanderMeer:163918, PhysRevLett.101.145002, doi:10.1063/1.1889444, PhysRevLett.103.194804, Couperus2017,Kirchen:456053,JalasPRL2021}), and the dynamics of the injected electrons in the resulting plasma fields~\cite{ASSMANN1998544,PhysRevE.74.026501,Ferran_Pousa2019-ew} dictate the final properties of the generated beams.
These processes can be controlled, up to a certain extent, by the parameters and design properties of the setup.
Typical examples include the plasma density profile (e.g.,~\cite{Kirchen:456053,JalasPRL2021,doi:10.1063/1.4958649,PhysRevSTAB.13.091301,PhysRevLett.110.185006,PhysRevLett.115.155002,doi:10.1063/1.4997606}), the properties of the laser pulse~\cite{PhysRevX.10.031039}, or the use of external laser guiding~\cite{PhysRevLett.122.084801, PhysRevX.12.031038, oubrerie2021controlled}.
Careful tuning and optimization of these parameters is critical for realizing LPAs that are capable of delivering the high beam quality and stability demanded by applications, particularly for free-electron lasers~\cite{Wang2021}, storage ring injectors~\cite{HILLENBRAND2014153,PhysRevAccelBeams.24.111301}, and future colliders~\cite{PhysRevSTAB.13.101301}.

Due to the complexity of the physical processes involved, the optimization of an LPA design requires the use of high-fidelity particle-in-cell (PIC) simulations~\cite{8659392} where the self-consistent interaction between particles and electromagnetic fields is computed with minimal assumptions.
However, the high computational cost associated with these simulations makes optimizing over a large set of parameters practically unfeasible.
This limits the number of configurations that can be explored for achieving optimal performance.

Developing more efficient techniques for optimizing the design of LPAs is therefore an important step toward realizing the full potential of these novel accelerators.
Besides the continued growth of available computing power, two approaches for more affordable optimization can be identified: reducing the number of simulations required to find the best-performing configuration, and reducing the cost per simulation.

The number of required simulations can be minimized by utilizing advanced algorithms that predict and evaluate only the most promising configurations throughout an optimization run.
An example of this is Bayesian optimization~\cite{Jones1998}, a machine learning-based technique that has gained popularity within the accelerator community~\cite{hanuka2019online, PhysRevLett.124.124801, ShallooNatComm2020, PhysRevAccelBeams.24.072802, Roussel2021, JalasPRL2021}.
This method generates a surrogate model of the simulation outcome (typically using Gaussian processes~\cite{RasmussenW06})  and suggests the most promising candidates for evaluation based on a balance between exploration (evaluating unmapped regions of the parameter space where new optima could be located) and exploitation (further sampling around known optima).
The underlying model is continuously updated with the results from new evaluations, allowing for more promising and accurate suggestions in successive iterations.
With this approach, the method can identify global optima with a reduced number of evaluations.

The computational cost of the individual simulations can be mitigated by making use of reduced models that sacrifice generality or accuracy by introducing physical approximations.
This can involve both reducing the dimensionality (e.g., assuming quasi-cylindrical symmetry~\cite{LIFSCHITZ20091803}) or neglecting certain physical properties of the laser-plasma interaction that are not dominant in the problem at hand.
Common examples of the latter include the use of a laser envelope model~\cite{Mora1997217} or assuming the wakefield to be quasi-static~\cite{Mora1997217,PhysRevA.41.4463}.
In principle, simulations with such reduced models can fully replace a complete PIC description if they accurately capture all the relevant physics involved.
In other cases, they provide an approximate solution from which useful information might still be extracted.

In this paper, we show that the computational cost of Bayesian optimization can be further reduced with the assistance of inexpensive reduced-model evaluations that are performed in tandem with costly high-fidelity simulations.
The inexpensive evaluations are used to dynamically probe regions of high interest and gather information that improves the predictions of the most promising configurations to evaluate at high fidelity.
This strategy is enabled by the use of a multitask Gaussian process model~\cite{NIPS2007_66368270, NIPS2013_f33ba15e, JMLR:v20:18-225}, whereby the correlation between the outputs of different tasks (i.e., the two levels of fidelity in the proposed method) is learned so that information gained on one task results in an improved model of the other.
In this way, the need for high-fidelity simulations is further reduced, leading to a faster and cheaper optimization.
This is demonstrated here by a proof-of-principle study combining the simulation codes \fbpic~\cite{LEHE201666} and \waket~\cite{Ferran_Pousa_2019}, which provide a full PIC description in quasi-3D geometry and inexpensive reduced models, respectively.

\section{MULTITASK BAYESIAN OPTIMIZATION}

Bayesian optimization is an efficient technique for the global optimization of black-box functions that are noisy and expensive to evaluate.
It operates by building a probabilistic \textit{surrogate model} of the \textit{objective function} $f$ (the function to minimize or maximize) that is cheaper to evaluate than $f$ and from which the most promising points to query can be determined by maximizing an \textit{acquisition function}.
The surrogate model is typically obtained by performing Gaussian process regression over the available data.
This provides an estimate of $f$ and its associated uncertainty at any point of the parameter space.
Determining which points to evaluate next depends on a balance between querying around known optima or exploring regions of high uncertainty where new optima could be identified.
This balance is quantified by the acquisition function, and the points that maximize it are deemed as most promising for future evaluation.
A typical choice for the acquisition function is the expected improvement~\cite{Jones1998}, which quantifies how much a new evaluation is expected to improve over the current optimum.
Once the new evaluations are completed, the Gaussian process model is updated with the obtained data and the same procedure is repeated.
This continuously improves the accuracy of the model and of the suggested evaluations.

With the use of a multitask Gaussian process (MTGP)~\cite{NIPS2007_66368270}, Bayesian optimization can be extended to a collection of objective functions $f_1,\ldots, f_N$ from $N$ different \textit{tasks}.
The MTGP learns the correlations between them and provides a surrogate model of each objective that features a reduced uncertainty by incorporating information from highly correlated tasks.
This exchange of information was originally proposed as a way of transferring the knowledge of previous optimizations to new tasks in order to optimize them more efficiently~\cite{NIPS2013_f33ba15e}.
Here, we make use of the approach described in Ref.~\cite{JMLR:v20:18-225}, where an inexpensive task $f_R$ (the reduced physical models) is used to assist in the optimization of a costly function $f_H$ (the high-fidelity PIC simulations) so that the number of required evaluations of $f_H$ is reduced.
This strategy is a special case of multi-fidelity optimization  where only two discrete levels of fidelity are considered.
Alternative multi-fidelity algorithms adapted to multiple objectives have also been explored in the context of particle accelerators~\cite{Irshad_arxiv_2021}.

In this two-task approach, the covariance function---or \textit{kernel}---that enables the MTGP to transfer information between tasks $t$ and $t'$ with inputs $x$ and $x'$ is defined as $k( (t,\mathbf{x}), (t',\mathbf{x}') ) = B_{tt'} \kappa(\mathbf{x},\mathbf{x}')$~\cite{JMLR:v20:18-225}.
This expression determines the covariance between data points from different tasks by assuming that both tasks share the same kernel  $\kappa(\mathbf{x},\mathbf{x}')$ for the input parameters (here, a Mat\'ern 5/2 kernel~\cite{RasmussenW06} is used) and that the task covariance can be captured separately by a $2 \times 2$ matrix $B$ where element $B_{tt'}$ is the covariance between $t$ and $t'$.
The coefficients of $B$ as well as the parameters of $\kappa(\mathbf{x},\mathbf{x}')$ are kernel hyperparameters that are inferred from the available data by maximizing marginal likelihood~\cite{JMLR:v20:18-225}.
The degree of inter-task correlation can be quantified by $\rho^2 = B_{tt'}^2/(B_{tt} B_{t't'})$, which ranges between $\rho^2=0$ (no correlation) to $\rho^2=1$ (maximum correlation).

Using this MTGP model, the algorithm performs a Bayesian optimization loop whereby batches of $n_R$ reduced-model simulations and $n_H$ high-fidelity simulations (with $n_H \leq n_R$) are executed in tandem.
At each iteration, the optimizer (i) fits an MTGP to the available data, (ii) determines a set $\{ \mathbf{x}_i \}_{i=1,\ldots,n_R}$ of the $n_R$ most promising points to query by maximizing noisy expected improvement~\cite{10.1214/18-BA1110} on the MTGP model for the \textit{high-fidelity} output $f_H(\mathbf{x})$, (iii) evaluates these $n_R$ points using \textit{reduced-model} simulations, (iv) updates the MTGP with the obtained results, (v) evaluates the $n_R$ points in the updated surrogate model of $f_H(\mathbf{x})$ to select the $n_H$ points with the most promising outcome, and (vi) evaluates the reduced sample of $n_H$ points with \textit{high-fidelity} simulations.
To start the optimization, initial samples of $n_{R,i}$ reduced-model simulations and $n_{H,i}$ high-fidelity simulations are generated by using two separate scrambled Sobol sequences~\cite{OWEN1998466} of input parameters.

This workflow has been implemented in a Python package that can take advantage of the capabilities of high-performance computing facilities.
The Bayesian optimization functionality is based on the \textsc{Ax} library~\cite{Bakshy2018AEAD} and can be executed on both CPUs and GPUs.
The allocation of computing resources for the optimizer and the simulations, as well as the coordination, execution, and communication between them, is orchestrated by the \textsc{libEnsemble} library~\cite{libEnsemble2022}.
This allows for the concurrent evaluation of multiple simulations that can make use of variable resources (number of CPUs and GPUs) across the available computing nodes.

\section{PROOF-OF-PRINCIPLE STUDY}


The effectiveness of the proposed algorithm is demonstrated here by a proof-of-principle optimization study combining the simulation codes \fbpic~\cite{LEHE201666} and \waket~\cite{Ferran_Pousa_2019}.
While \fbpic\ provides a high-fidelity, fully electromagnetic PIC description of the LPA physics in quasi-3D geometry~\cite{LIFSCHITZ20091803}, \waket\ allows for inexpensive simulations by using a reduced quasi-static wakefield model with 2D cylindrical symmetry~\cite{PhysRevAccelBeams.21.071301} and a laser envelope model~\cite{Benedetti_2017}.

The setup to be optimized is an LPA booster stage for an externally injected electron bunch.
Given a fixed laser driver and plasma profile, the goal is to determine the bunch current profile that results in the lowest energy spread with the highest possible charge and energy.
This involves optimizing the net beam loading effect~\cite{vanderMeer:163918, PhysRevLett.101.145002, doi:10.1063/1.1889444, JalasPRL2021, Kirchen:456053} throughout the LPA, a nontrivial process affected by laser dephasing, depletion, and diffraction for which no analytical theory is available and that must therefore be addressed with simulations.

To simultaneously achieve low energy spread, high charge, and high energy, these quantities are combined into a single objective to maximize:
\begin{equation}\label{eq:objective_function}
    f = \frac{k_Q E_\mathrm{MED}\mathrm{[GeV]}}{k_\mathrm{MAD}} \ ,
\end{equation}
where $k_Q = Q_\mathrm{tot} /Q_\mathrm{ref}$ is the ratio between the total $Q_\mathrm{tot}$ and a reference  $Q_\mathrm{ref}=\SI{10}{\pico\coulomb}$ charge, $k_\mathrm{MAD} = \Delta E_\mathrm{MAD} / \Delta E_\mathrm{MAD, ref}$ is the ratio between the relative energy spread $\Delta E_\mathrm{MAD}$ and a reference value $\Delta E_\mathrm{MAD, ref} = 10^{-2}$, and $E_\mathrm{MED}$ is the median energy.
The use of the median absolute deviation (MAD) energy spread and median energy provides a robust characterization of the energy spectrum in distributions with outliers, as typically observed in LPAs~\cite{JalasPRL2021, Kirchen:456053}.
The value of $f$ given by Eq. (\ref{eq:objective_function}) can span over several orders of magnitude and feature sharp extremes that are not ideal for Gaussian process modeling.
To alleviate this, the objective is internally treated by the optimizer as $\log(f)$.

The parameters of the laser driver are an energy $E_L=\SI{10}{\joule}$, an FWHM duration $\tau_\mathrm{FWHM}=\SI{25}{\femto\second}$, a focal spot size $w_0=\SI{40}{\micro\metre}$, a wavelength $\lambda_0=\SI{800}{\nano\metre}$, and a peak normalized vector potential $a_0\simeq2.6$.
The plasma density profile is a simple \SI{10}{\centi\meter}-long flat-top with an on-axis electron density $n_{e,0}=\SI{2e17}{\per\cubic\centi\metre}$ and a parabolic radial profile for laser guiding $n_e(r) = n_{e,0} + r^2 / (\pi  r_e  w_0^4)$~\cite{PhysRevLett.72.2887}.
The externally injected electron bunch has an initial energy $E_{b,0}=\SI{200}{\mega\electronvolt}$ with an rms energy spread of \SI{0.1}{\%}.
It features a normalized emittance of $\epsilon_{n,x}=\SI{3}{\micro\metre}$ in the horizontal direction and of $\epsilon_{n,y}=\SI{0.5}{\micro\metre}$ in the vertical plane.
This difference between the $x$ and $y$ emittances typically arises in LPAs based on ionization injection as a result of the laser polarization~\cite{JalasPRL2021}.
Here, it is included in the externally injected bunch in order to ensure a bias between the two simulation codes, as this asymmetry can only be fully captured by the high-fidelity \fbpic\ simulations.
The initial transverse size is matched to the focusing strength in the plasma, allowing for emittance preservation~\cite{ASSMANN1998544, PhysRevSTAB.15.111303}.
The longitudinal profile of the bunch is assumed to be trapezoidal, as it is known to be well suited for beam loading~\cite{PhysRevLett.101.145002}, and features smooth Gaussian ramps ($\SI{1}{\micro\metre}$ rms) at the head and tail.
The parameters exposed to the optimizer are the current at the head $I_\mathrm{h}$ and tail $I_\mathrm{t}$, the bunch length $L_b$, and its longitudinal position in the wake, parameterized by the distance $\Delta z_\mathrm{l,h} = z_l - z_\mathrm{h}$ between the head of the bunch $z_\mathrm{h}$ and the center of the laser driver $z_l$.
They are allowed to vary in the following ranges: $I_h \in [0.1, 10] \, \si{\kilo\ampere}$, $I_t \in [0.1, 10] \, \si{\kilo\ampere}$, $L_b \in [1, 20] \, \si{\micro\metre}$, and $\Delta z_{l, h} \in [40, 60] \, \si{\micro\metre}$.

\begin{figure}
    \includegraphics[width=\columnwidth]{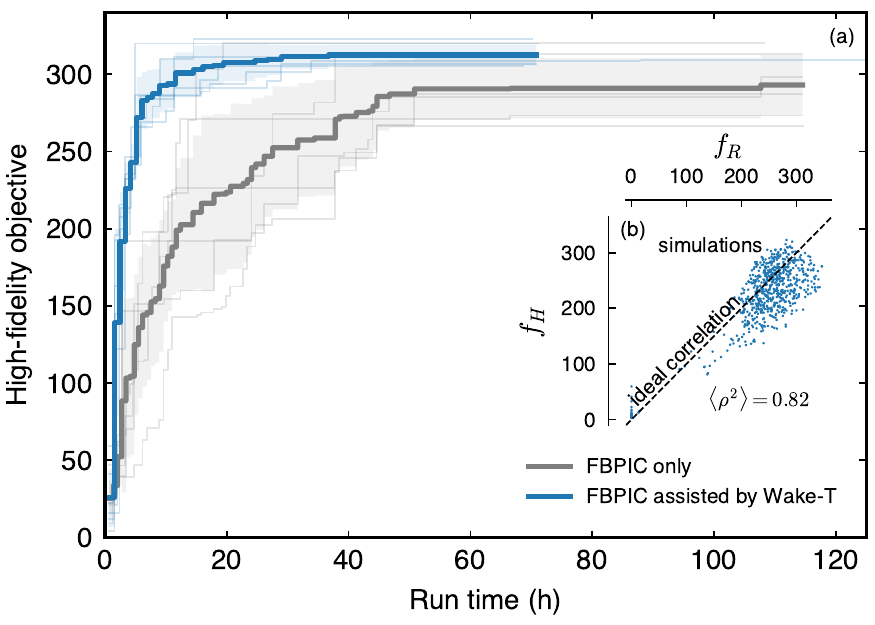}%
	\caption{\label{fig:mt_vs_sf} (a) Average (thick line) and standard deviation (shaded area) of the evolution of the high-fidelity \fbpic\ objective with and without the assistance of reduced-model simulations with \waket. Six runs (thin lines) were performed for each case. (b) Correlation between the \fbpic\ ($f_H$) and \waket\ ($f_R$) objectives in the multitask optimization as obtained from individual simulations.}
\end{figure}

The \fbpic\ simulations are performed using the boosted frame technique~\cite{PhysRevLett.98.130405, Vay2011} with a Lorentz boost factor of 25.
The longitudinal and radial resolutions are $dz=\lambda_0/80$ and $dr=k_p^{-1}/20$, respectively, where $k_p=(n_{e,0} e^2/m_e \epsilon_0c^2)^{1/2}$ is the plasma wave number, $e$ is the elementary charge, $\epsilon_0$ is the vacuum permittivity and $c$ is the speed of light. The number of particles per cell is $2$ in both $z$ and $r$, and $8$ in the azimuthal direction.
Three azimuthal modes are used to properly describe the ellipticity of the electron bunch.
The simulations are performed on a single NVIDIA A100 GPU and have a typical execution time of $\SI{\sim 40}{\minute}$.
The \waket\ simulations have a resolution of $dz=c\tau_\mathrm{FWHM}/40$ and $dr=k_p^{-1}/20$ with 2 particles per cell.
Each simulation is performed on a single core of an AMD EPYC 7643 CPU, with a typical execution time of $\SI{\sim 3}{\minute}$.
The entire optimization is carried out in one compute node with 96 CPU cores and 4 GPUs, one of which is allocated for the optimizer.
With this setup, batches of either $n_R = 96$ concurrent \waket\ simulations or $n_H=3$ concurrent \fbpic\ simulations can be performed.

To quantify the performance gain from the multitask approach, the same physical setup is also optimized solely with batches of 3 \fbpic\ simulations using a Bayesian algorithm based on a standard single-task Gaussian process model.
This optimization is carried out in the same hardware and uses the same initialization routine and acquisition function as the multitask case.
Since each optimization run evolves differently---both the initial sample of points and the optimization of the acquisition function include a certain degree of randomness---a total of 6 independent multitask and single-task optimizations have been carried out to determine the average evolution and its variance.

The results of this scan, shown in Fig.~\ref{fig:mt_vs_sf}, indicate that incorporating reduced-model simulations into the optimization leads, on average, to an order-of-magnitude speedup in terms of the time to converge to a solution as well as a reduced variability in the convergence rate.
For example, an average objective value of $f_H=280$ is reached after $\SI{\sim 6}{\hour}$ when the optimization is assisted with \waket\ simulations, while this number grows to $\SI{\sim 45}{\hour}$ when only \fbpic\ is used.
This boost in performance is achieved despite the outcome of both codes not being in full agreement with each other, as evidenced in Fig.~\ref{fig:mt_vs_sf}(b).
However, owing to the high degree of correlation between them ($\langle\rho^2\rangle \simeq 0.82$, where $\langle \rangle$ denotes average over the 6 runs), the multitask algorithm can capture the bias of the reduced model with respect to the high-fidelity simulations and extract useful information from it.

\begin{figure}
    \includegraphics[width=\columnwidth]{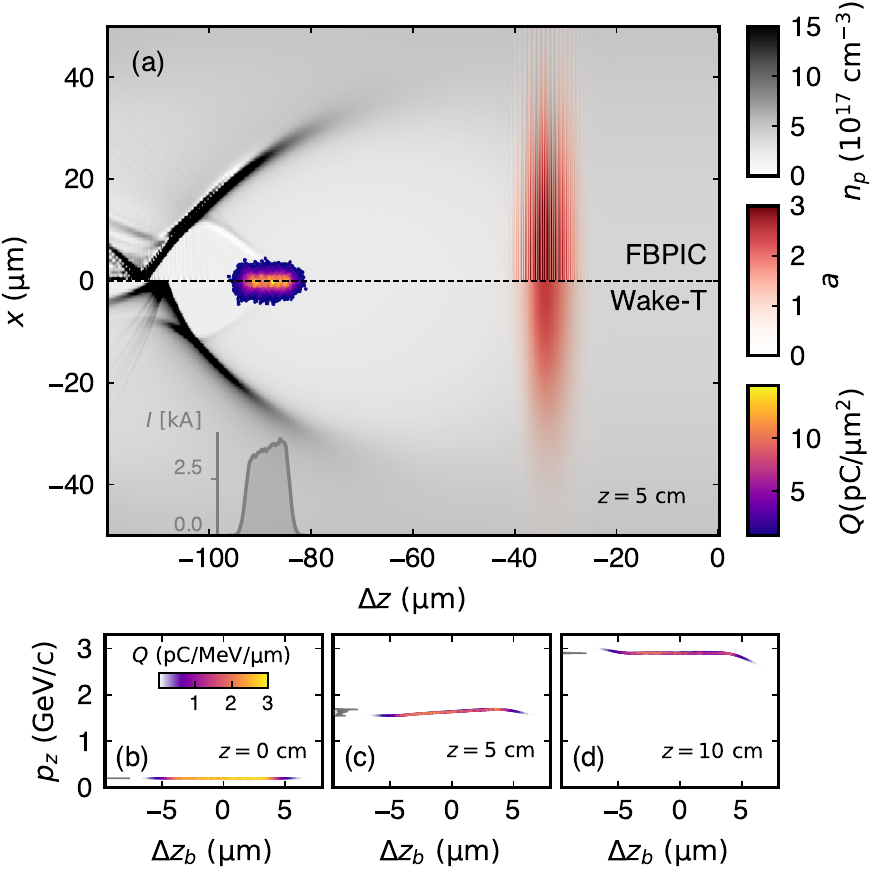}%
	\caption{\label{fig:wakefields} Outcome of the highest scoring \fbpic\ simulation. (a) Plasma wakefields at the center of the LPA as obtained from \fbpic\ (top) and \waket\ (bottom). Longitudinal phase space at the (b) start, (c) middle, and (d) end of the \fbpic\ simulation. $\Delta z$ and $\Delta z_b$ are the longitudinal positions relative to the front of the simulation box and the beam center, respectively.}
\end{figure}

This approach successfully manages to optimize the given setup.
The highest scoring \fbpic\ simulation ($f_H\simeq323$) from the 6 multitask optimizations corresponds to a configuration with $I_h=\SI{4.26}{\kilo\ampere}$, $I_t=\SI{3.50}{\kilo\ampere}$, $L_b=\SI{6.35}{\micro\metre}$ and $\Delta z_{l,h}=\SI{55.2}{\micro\metre}$, which results in a total charge of \SI{114.6}{\pico\coulomb}, a mean energy of \SI{2.9}{\giga\electronvolt}, and a relative energy spread of \SI{0.1}{\%} (MAD).
Fig. \ref{fig:wakefields} shows the outcome of the \fbpic\ and \waket\ simulations for this working point.
Certain differences can be observed in the plasma wake, particularly towards the back, where highly relativistic plasma electrons cannot be accurately modeled within the quasi-static approximation.
The evolution of the longitudinal phase space seen in Figs. \ref{fig:wakefields}(b)-(d) shows that, as originally intended, an optimal net beamloading is achieved at the end of the LPA.
Even though the energy spread can be locally high at some points during acceleration, the laser evolution and the subsequent changes to the plasma wake along the LPA end up resulting in a flattened energy distribution.

\begin{figure}
    \includegraphics[width=\columnwidth]{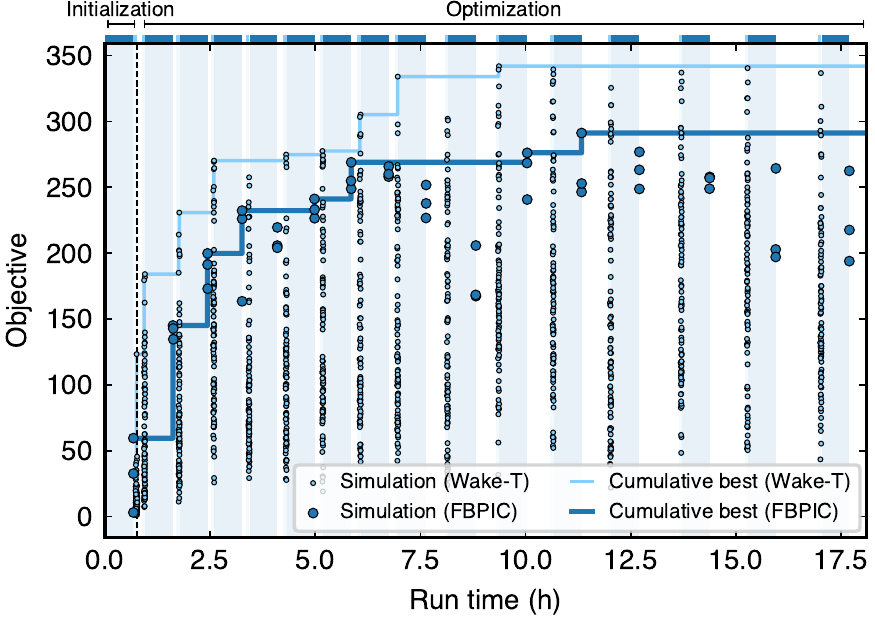}%
	\caption{\label{fig:mt_details} Evolution of a multitask optimization with alternating batches of \waket\ and \fbpic\ simulations, indicated by the shaded area. Intervals between batches correspond to the time when the optimizer is computing the next set of configurations to evaluate. Outcomes of each simulation and the evolution of the cumulative best objective are also included.}
\end{figure}

A detailed view of the sequence of simulation batches and the evolution of $f_R$ and $f_H$ in a multitask optimization is included in Fig. ~\ref{fig:mt_details}.
The \waket\ simulations have a negligible cost compared to the \fbpic\ batches, allowing for broad and inexpensive exploration so that only the most promising configurations are evaluated at high fidelity.
This allows for a much faster convergence of $f_H$, which evolves at virtually the same rate as $f_R$ despite the reduced number of simulations.
However, one potential drawback of performing large batches of reduced-model simulations is an increased cost of suggesting new configurations (i.e., of fitting the MTGP and optimizing the acquisition function) due to the rapid growth in the total number of evaluations.
This is evidenced in Fig. ~\ref{fig:mt_details}, where the intervals between simulation batches progressively widen as the total number of evaluations increases.
Therefore, determining an adequate ratio between $n_R$ and $n_H$ is of high relevance for a well-performing optimization.
Otherwise, the cost savings from the increased convergence rate could be counterbalanced, at least in part, by the growing cost of the multitask optimizer.

\begin{figure}
    \includegraphics[width=\columnwidth]{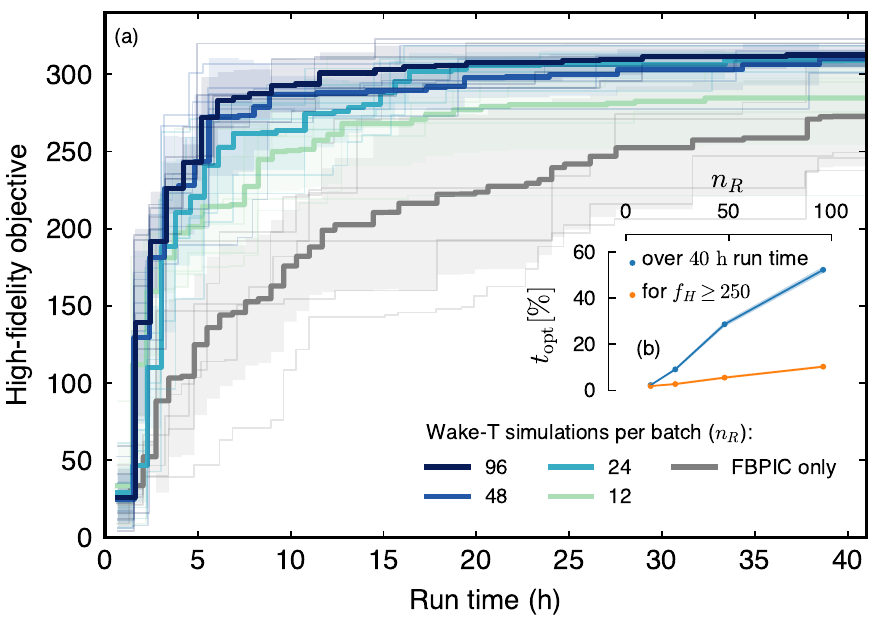}%
	\caption{\label{fig:lofi_hifi_ratio} (a) Evolution of the high-fidelity objective in multitask optimizations using a different number of \waket\ simulations per batch compared against a single-task (\fbpic\ only) benchmark. (b) Percentage of total run time consumed by the optimizer as a function of $n_R$. This percentage is measured both over the $\SI{40}{\hour}$ period shown and, alternatively, over the time needed to reach an objective $f_H \geq 250$.}
\end{figure}

The influence of the ratio between $n_R$ and $n_H$ is investigated here with a series of optimizations where the  number of \waket\ simulations per batch is varied.
In addition to the original study with $n_R=96$, three more cases (each of them consisting of 6 independent runs) with $n_R=\SIlist{48;24;12}{}$ are included.
For each case, the evolution of $f_H$ as well as the fraction of time that is spent purely in the optimizer, $t_\mathrm{opt}$, are quantified.
The general outcome of this scan is that reducing $n_R$ leads to a slower convergence in terms of the number of iterations but to a faster optimizer (i.e., smaller $t_\mathrm{opt}$).
These two effects partially compensate each other in terms of total run time, leading to no significant differences between the cases with $n_R=\SIlist{96;48;24}{}$, as seen in Fig. \ref{fig:lofi_hifi_ratio}.
To achieve an objective $f_H \geq 250$, which is reached by all runs, the time consumed by the optimizer is moderate in all cases, ranging from $t_\mathrm{opt}\simeq\SI{10}{\%}$ when $n_R=96$ to $t_\mathrm{opt}\simeq\SI{2}{\%}$ when $n_R=12$.
However, when quantifying $t_\mathrm{opt}$ over the $\SI{40}{\hour}$ period displayed in Fig. \ref{fig:lofi_hifi_ratio}, the optimizer becomes the dominant contribution to the total run time ($t_\mathrm{opt}\simeq\SI{52}{\%}$) when $n_R=96$ while remaining negligible ($t_\mathrm{opt}\simeq\SI{2}{\%}$) when $n_R=12$.
This is not particularly concerning here, as the case with $n_R=96$ reaches a close-to-optimal objective well before the $\SI{40}{\hour}$ threshold.
However, if a higher number of iterations were required, such as in a case where the reduced models and the high-fidelity simulations are not as well correlated, it could lead to a significant loss in performance.
Based on the results from this scan, $n_R=24$ appears to be an adequate choice that leads to virtually the same rate of convergence as cases with higher $n_R$ while allowing, if needed, for a larger number of iterations.

\begin{figure}
    \includegraphics[width=\columnwidth]{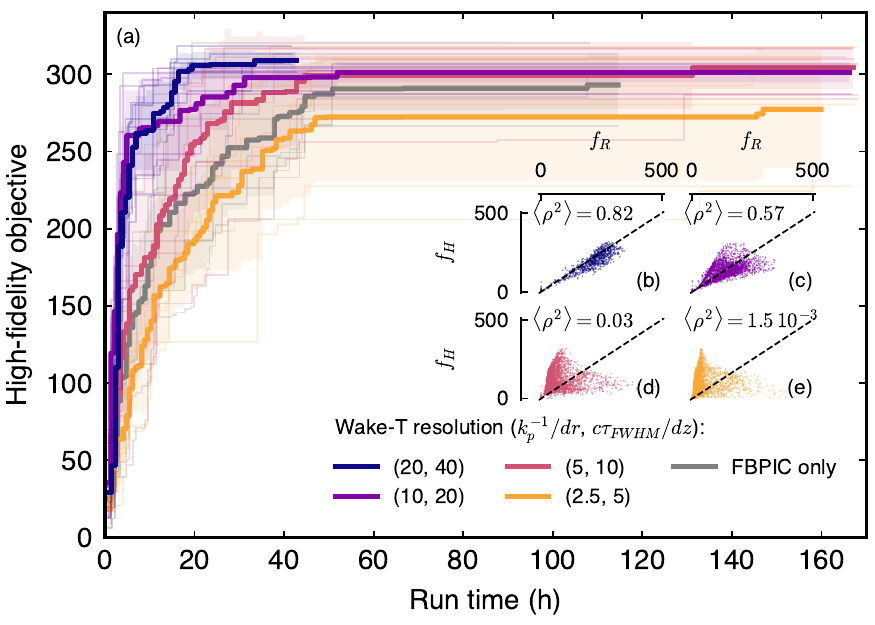}%
	\caption{\label{fig:lofi_quality} (a) Evolution of the high-fidelity objective in multitask optimizations assisted by \waket\ simulations of different resolutions compared against a single-task (\fbpic\ only) benchmark. (b), (c), (d) and (e) show the correlation between \waket\ ($f_R$) and \fbpic\ ($f_H$) results in each case.}
\end{figure}

In general, depending on the physical problem to optimize, different reduced models of varying fidelity and cost might be available.
As such, studying the behavior of the multitask method under varying degrees of inter-task correlation is of high relevance for its general applicability.
In particular, it is important to ensure that the method converges to a meaningful optimum despite any degradation of the information gained from the reduced model and a potential increase in overall costs.
To test this, an additional set of optimizations has been performed where the fidelity of $f_R$ is reduced by decreasing the resolution of the \waket\ simulations.
In addition to the original setup, 3 cases with a factor of $\SIlist{2;4;8}{}$ lower resolution in both $z$ and $r$ are included.
Due to the expected reduction in convergence rate, all optimizations are performed with $n_R=24$ to allow for a larger number of iterations without significantly increasing $t_\mathrm{opt}$.
The results from this study, summarized in Fig. \ref{fig:lofi_quality}, clearly indicate that a loss in correlation directly translates into a slower convergence rate.
However, even with a moderate correlation ($\langle\rho^2\rangle\simeq0.57$), the multitask algorithm can still provide a significant performance gain.
Only when the two tasks are essentially independent (i.e., $\langle\rho^2\rangle\sim 10^{-3}$ in the lowest resolution case) does the rate of convergence decrease below the single-task benchmark.
This is because even though the MTGP reduces to a single task when $\rho^2=0$~\cite{JMLR:v20:18-225}, the inaccurate data from $f_R$ can still influence the surrogate model of $f_H$ until sufficient evaluations to infer the lack of correlation have been gathered.
As such, the multitask technique converges towards the optimum even with unreliable or misleading reduced-model data, and provides a performance boost over single-task optimization as long as a meaningful inter-task correlation can be recognized.

\section{CONCLUSION}

The proposed multitask method introduces the capability of leveraging reduced physical models for assisting in the Bayesian optimization of LPAs and lowering the need for costly high-fidelity simulations.
In a proof-of-principle study combining the simulation codes \fbpic\ (high fidelity) and \waket\ (reduced models), this technique demonstrates an order-of-magnitude speedup over an equivalent single-task Bayesian optimization consisting solely of \fbpic\ simulations.
This improvement in performance depends on the ratio of reduced-model to high-fidelity simulations, the cost difference between them, and their degree of correlation.
An excessive number of reduced model simulations can increase the computational cost of suggesting new configurations, thus partially counterbalancing the gain in performance, while carrying out too few can slow down the convergence rate.
Batches of $n_R=24$ \waket\ simulations and $n_H=3$ \fbpic\ simulations were found to be an adequate balance in the presented study.
The choice of a reduced model that correlates well with the high-fidelity simulations is essential for achieving a significant speedup, although the algorithm converges towards the optimum even if no information is gained from the inexpensive simulations.
The high computational efficiency of this method allows for the cost-effective optimization of LPAs in large parameter spaces.
This is a critical step toward unlocking the full potential of these devices and fulfilling the high beam quality requirements of applications such as free-electron lasers, storage-ring injectors, and future particle colliders.

\begin{acknowledgments}
This material is based upon work supported by the U.S. Department of Energy, Office of Science, under contract numbers DE-AC02-06CH11357 and DE-AC02-05CH11231 and by the Exascale Computing Project (17-SC-20-SC).
This research was supported in part through the Maxwell computational resources operated at DESY, Hamburg, Germany.
\end{acknowledgments}

\appendix


\bibliography{mtbo}

\end{document}